\documentclass[%
 reprint,
 amsmath,amssymb,
 aps,
]{revtex4-2}

\usepackage{hyperref}
\usepackage{placeins}
\usepackage{graphicx}
\usepackage{dcolumn}
\usepackage{bm}

\begin{document}

\preprint{APS/123-QED}

\title{Separating non-collective effects in d-Au collisions\\}

\author{Satya Ranjan Nayak$^{1}$}
 \email{satyanayak@bhu.ac.in}
\author{Akash Das$^{2}$}
  \email{24pnpo01@iiitdmj.ac.in}

\author{B. K. Singh$^{1,2}$}%
 \email{bksingh@bhu.ac.in}
 \email{director@iiitdmj.ac.in (Corresponding author)}
\affiliation{$^{1}$Department of Physics, Institute of Science,\\ Banaras Hindu University (BHU), Varanasi, 221005, INDIA. \\
$^{2}$Discipline of Natural Sciences, PDPM Indian Institute of Information Technology Design \& Manufacturing, Jabalpur, 482005, INDIA\\
}
\date{\today}

\date{\today}
\begin{abstract}

In this work, we present the multiplicity and yield of charged hadrons and particle ratios in d-Au collisions at $\sqrt{s_{NN}}= 200$ GeV using the PYTHIA8/Angantyr. The model reproduces the multiplicity ($N_{ch}$) and pseudo-rapidity distribution reasonably well in minimum-biased d-Au collisions without assuming the formation of a thermalized medium. The invariant yield from Angantyr underpredicts the data in central collisions. We discussed the similarity between the nuclear modification factor $R_{AA}$ and data/MC from Angantyr and the possibility of using it as a model-dependent observable to check in-medium effects. The data/MC suggests that the central d-Au collisions exhibit signals like baryon enhancement, but no high $p_T$ suppression was found. The d-Au collisions have a much smaller invariant yield than the thermal model calculation for similar $N_{part}$.

\end{abstract}

\maketitle


\section{\label{sec:level1}Introduction\protect\\ }
The quark-gluon plasma is a deconfined state of strongly interacting partons similar to the matter produced shortly after the ``Big Bang" \cite{Collins:1974ky,Cabibbo:1975ig,Chapline:1976gy}. Collider facilities such as BNL RHIC and CERN LHC often collide Heavy ion beams at ultra-relativistic speeds to achieve such a state. These experiments have been running for over two decades, yet we don't fully understand the limiting conditions to form a system that can be classified as a QGP without any debate. A QGP is identified by its signatures \cite{singh1993signals}, such as jet quenching \cite{PHENIX:2001hpc}, anisotropic flow \cite{Wiedemann:1997cr}, strangeness enhancement \cite{Rafelski:1982pu,Rafelski:1982ii}, baryon enhancement \cite{PHENIX:2023kax,PHENIX:2003iij,PHENIX:2001vgc}, etc. However, recent developments have shown that rope hadronization and string shoving \cite{Bierlich:2016vgw} can reproduce signals like strangeness enhancement \cite{Bierlich:2022ned,Bierlich:2017vhg}. Similarly, mechanisms like color reconnection \cite{Christiansen:2015yqa,Lonnblad:2023stc} can mimic the collective effect of QGP and give non-zero anisotropic flow. Hence, QGP-like effects arising from different mechanisms should be studied in detail before drawing any conclusion about the formation of QGP. These effects are crucial while studying systems like d-Au and He-Au collisions, which show signals of QGP despite extremely small system size. If a QGP is forming in such a collision, the thermal model calculations should describe the data. If the data can be described by a ``non-QGP" model, the collision is not expected to form QGP. 

The major goal of this work is to distinguish the non-collective aspects of a d-Au collision from the collective ones. The Angantyr model \cite{Bierlich:2018xfw} provides just the right tool for this study. The model simulates a heavy ion event based on the interactions of the wounded nucleons without assuming the formation of a QGP. Furthermore, a newly added mechanism known as the spatially constrained color reconnection (SC CR) \cite{Lonnblad:2023stc} can enhance the baryon production by introducing string junctions that can later hadronize into baryons. We will study the effect of SC CR in the d-Au collisions and its effect on the ``baryon enhancement" signal.

\begin{table*}
\caption{\label{tab:table1}%
The values of parameters in different PYTHIA tunes.
}
\begin{ruledtabular}
\begin{tabular}{ccccc}
\textrm{Parameter}&
\textrm{Monash}&
\textrm{QCD CR}&
\textrm{SC CR tune}&
\textrm{This work}\\
\colrule
MultiPartonInteractions:PT0Ref & 2.28 & 2.12 & 2.37 &  2.39 \\
HIMultiPartonInteractions:PT0Ref & 2.28 & 2.12 & 2.37 &  2.39 \\
StringFlav:ProbStoUD & 0.217 & 0.2 & 0.2 &  0.3 \\
StringFlav:probQQtoQ & 0.081 & 0.078 & 0.078 &  0.11 \\
BeamRemnants:primordialKTsoft  & 0.9 & 0.9 & 0.9 & 1.55 \\
BeamRemnants:primordialKThard  & 1.8 & 1.8 & 1.8 & 2.0 \\

\end{tabular}
\end{ruledtabular}
\end{table*}

\section{Angantyr}
PYTHIA is one of the most widely used all-purpose event generators that can be used for several collision systems \cite{Bierlich:2022pfr,Sjostrand:2004ef}. The Angantyr model was recently introduced by PYTHIA for the p-A and A-A collisions. The simulation starts by separating the individual NN sub-collisions based on a modified Glauber model. Unlike the traditional Glauber model, the current version considers the color fluctuations in the nucleon substructure (Glauber-Gribov color fluctuations) \cite{Alvioli:2013vk,Alvioli:2017wou}. After the sub-collisions are selected, the absorptive sub-collisions (where neither the target nor the projectile nucleon has interacted before) are treated as standard Non-Diffractive collisions \cite{Bierlich:2018xfw}. Once a nucleon is absorptively wounded, it can still interact with other target/projectile nucleons. Such sub-collisions are marked as secondary non-diffractive (wounded target/wounded projectile) and treated similarly to a Non-Diffractive collision between a proton and a pomeron ( here, the pomeron is treated as a physical object with parton densities). The double diffractive and elastic sub-collisions are treated accordingly, and then all sub-collisions are combined together into a heavy-ion event.

The PYTHIA models the soft particle production based on the multi-parton interaction (MPI) machinery \cite{Sjostrand:1987su}. It comes with a unique challenge: each MPI increases the number of long strings in the central rapidity region, leading to higher multiplicities overshooting the data. At the same time, it decreases the $<p_T>$ of the event. This problem is resolved by a mechanism called the Color Reconnection (CR) \cite{Sjostrand:1987su}, which refers to recombining the colored strings from neighboring $q\bar{q}$ dipoles to reduce the length of the strings and avoid additional particle production. The default MPI-based CR works reasonably well in the p-p collisions, but certain modifications are necessary to apply the CR mechanism in Angantyr. The current version is an extension of QCD CR where the strings are assigned color according to SU(3) color algebra instead of just the leading color \cite{Christiansen:2015yqa}. The default color reconnection (mode 0) in Angantyr performs color reconnections in individual sub-collisions independently before the hadronization stage. This is based on the assumption that strings from separate sub-collisions don't interact with each other. However, the strings from different sub-collisions of a heavy ion event can interact with each other. Hence, the color reconnections should be performed for the entire event. Performing color reconnections for the whole of the event comes with its own challenge, i.e, the spatial span of a heavy-ion event (about the sum of the diameters of two nuclei) is larger than the range of strong interactions. Hence, the color reconnections should be constrained spatially. A spatial constraint will allow color reconnections among nearby dipoles regardless of sub-collision. A detailed description along with the exact values of different parameters can be found in the corresponding papers \cite{Christiansen:2015yqa,Lonnblad:2023stc}. The parameters in ref. ~\cite{Lonnblad:2023stc} were tuned to p-Pb and Pb-Pb collision data at LHC energies. The tune in ref.~\cite{Lonnblad:2023stc} can't fully describe the identified particle spectra in pp collisions at 200 GeV. We modified a few parameters to better describe the identified particle spectra in pp collisions at 200 GeV. The mid-$p_T$ yield of all primary hadrons was smaller than the experimental data in the default tune. The value of BeamRemnants:primodialKTsoft and BeamRemnants:primodialKThard were increased to improve the qualitative description of identified particle spectra at mid-$p_T$. Similarly, the values of StringFlav:ProbStoUD and StringFlav:ProbQQtoQ were increased to enhance the yield of kaons and protons, respectively. In Fig. 1, we have shown the $p_T$ distribution of identified hadrons in pp collisions using both the current PYTHIA tune and the default tune for reference. The PYTHIA calculations provide a reasonable description of the invariant yields of pions, protons, and kaons. The values of the modified parameters and their values in different tunes can be found in Table 1. The $p/\pi^+$ and $k^+/\pi^+$ ratios deviate from the data in PYTHIA calculations. PYTHIA lacks mechanisms like partonic coalescence, which are essential for describing $p/\pi$ ratios \cite{Zhao:2021vmu}. The ratios obtained using PYTHIA can not be used as a reliable baseline. We want the reader to note these limitations before proceeding to the results section.


\begin{figure*}
\includegraphics[width=1\textwidth]{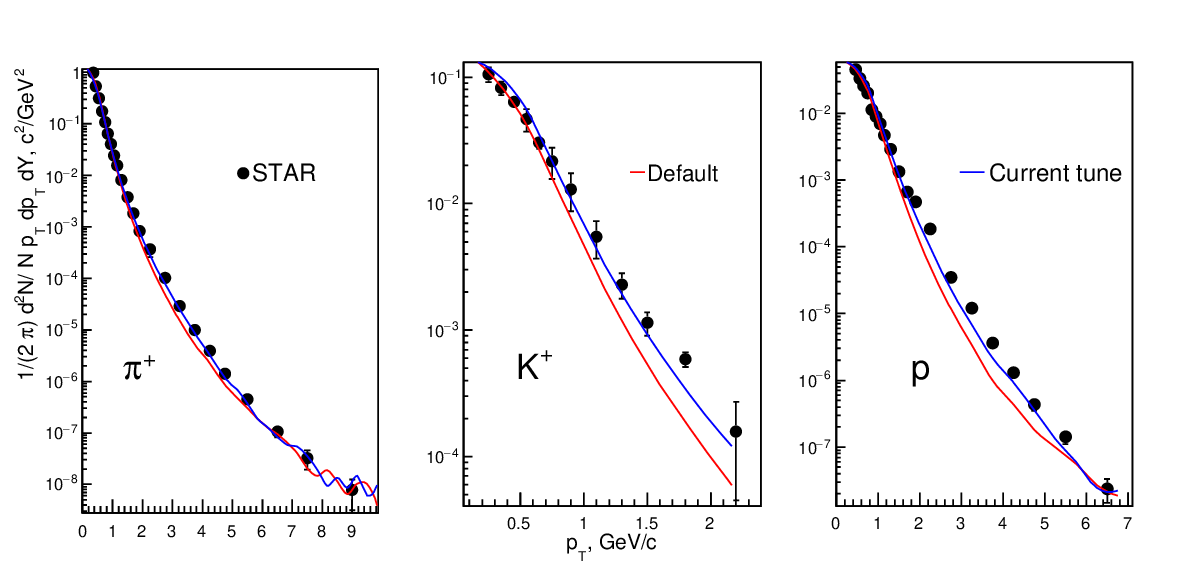}
\caption{\label{fig:3} The transverse momentum distribution of identified hadrons in pp collisions at 200 GeV.}
\end{figure*}

\begin{figure}[h]
\includegraphics[width=.5\textwidth]{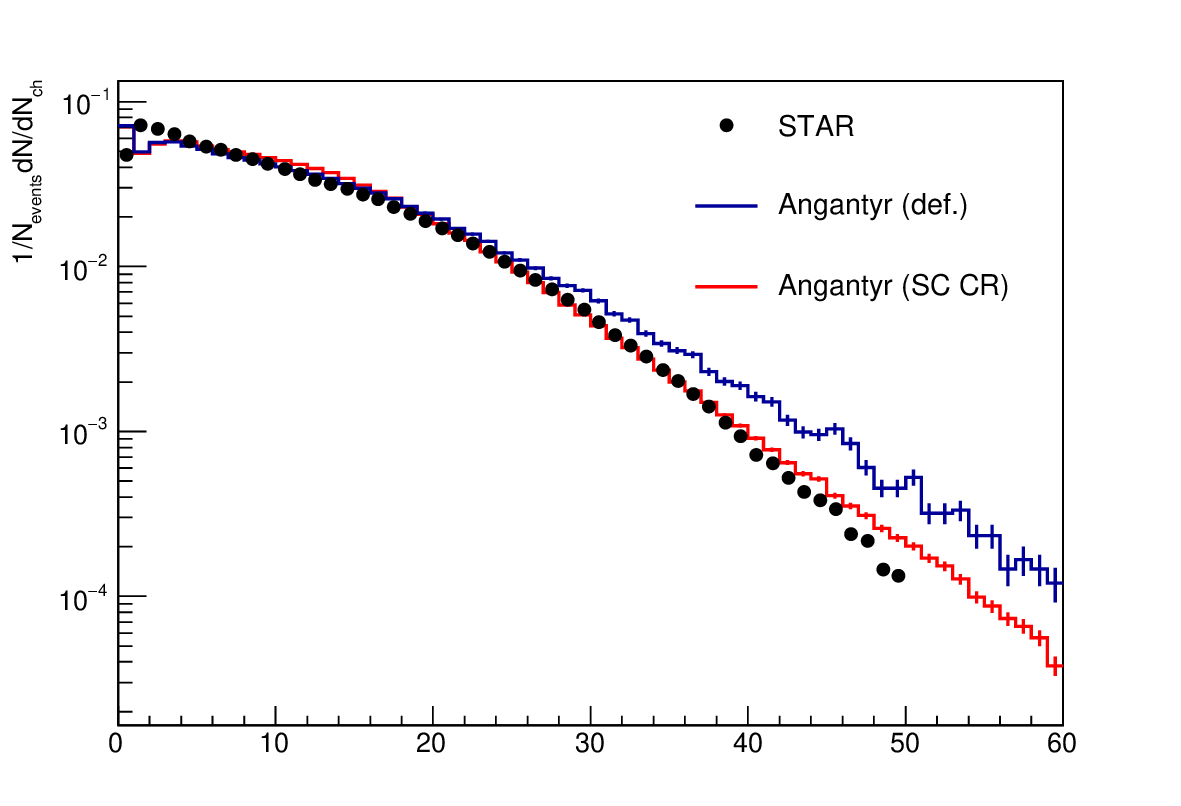}
\caption{\label{fig:3} The charged hadron multiplicity distributions in a minimum-biased d-Au collisions at 200 GeV. The data points are from STAR \cite{STAR:2006kxj}.}
\end{figure}

\begin{figure}[h]
\includegraphics[width=.5\textwidth]{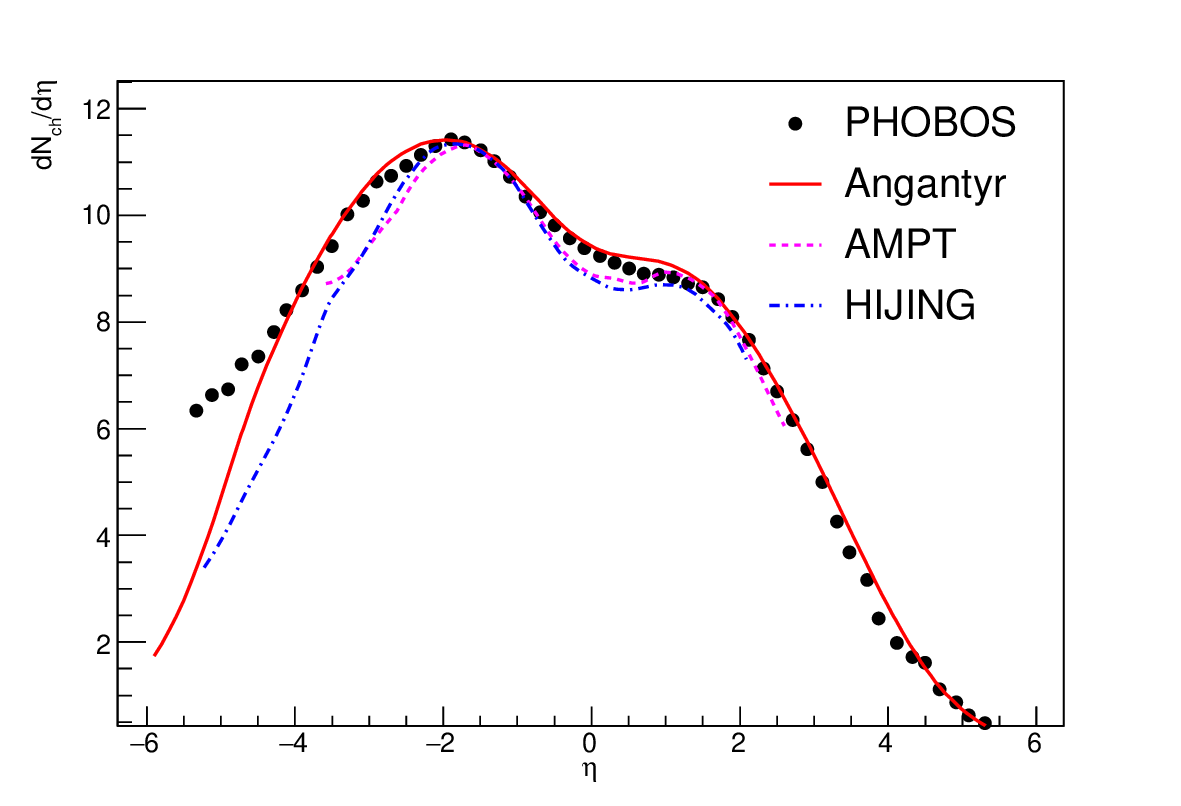}
\caption{\label{fig:3} Pseudo-rapidity distribution of charged hadrons in minimum biased d-Au collisions. The details about lines and markers can be found in the legend. The data points are from PHOBOS \cite{PHOBOS:2003fjw}.}
\end{figure}

\begin{figure*}
\includegraphics[width=1.0\textwidth]{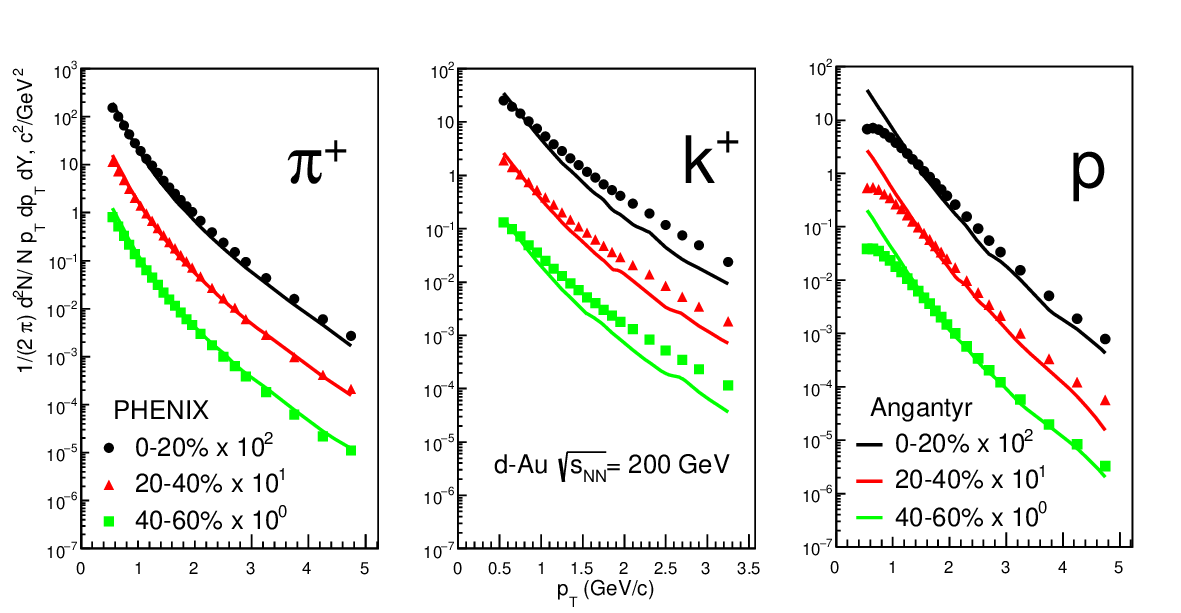}
\caption{\label{fig:8}Transverse momentum spectra of pions, kaons, and protons for different centralities in d-Au collisions at $\sqrt{s_{NN}}$= 200 GeV. Markers represent experimental data \cite{PHENIX:2013kod}, and the Angantyr results are shown as lines.}
\end{figure*}

\begin{figure}[h]
\includegraphics[width=.5\textwidth]{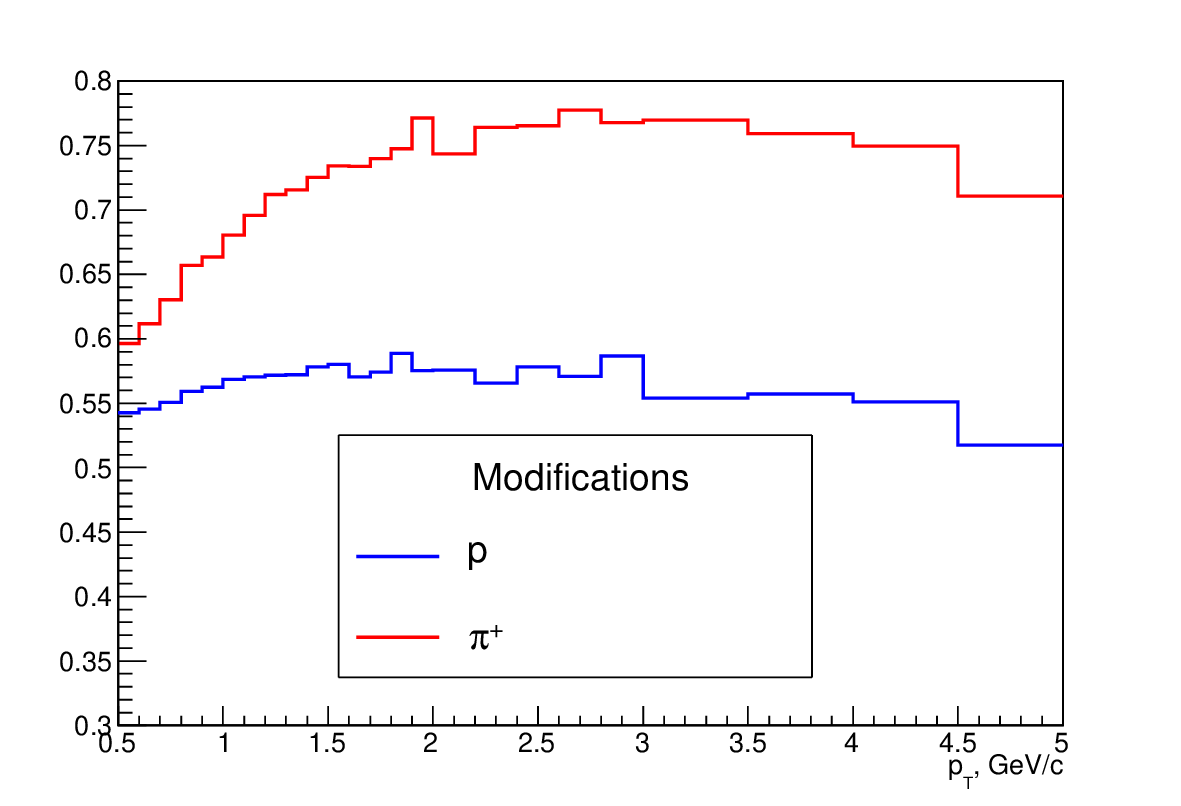}
\caption{\label{fig:3} The modification due to separation of sub-collisions in d-Au collisions at 200 GeV. }
\end{figure}
\begin{figure}[h]
\includegraphics[width=.5\textwidth]{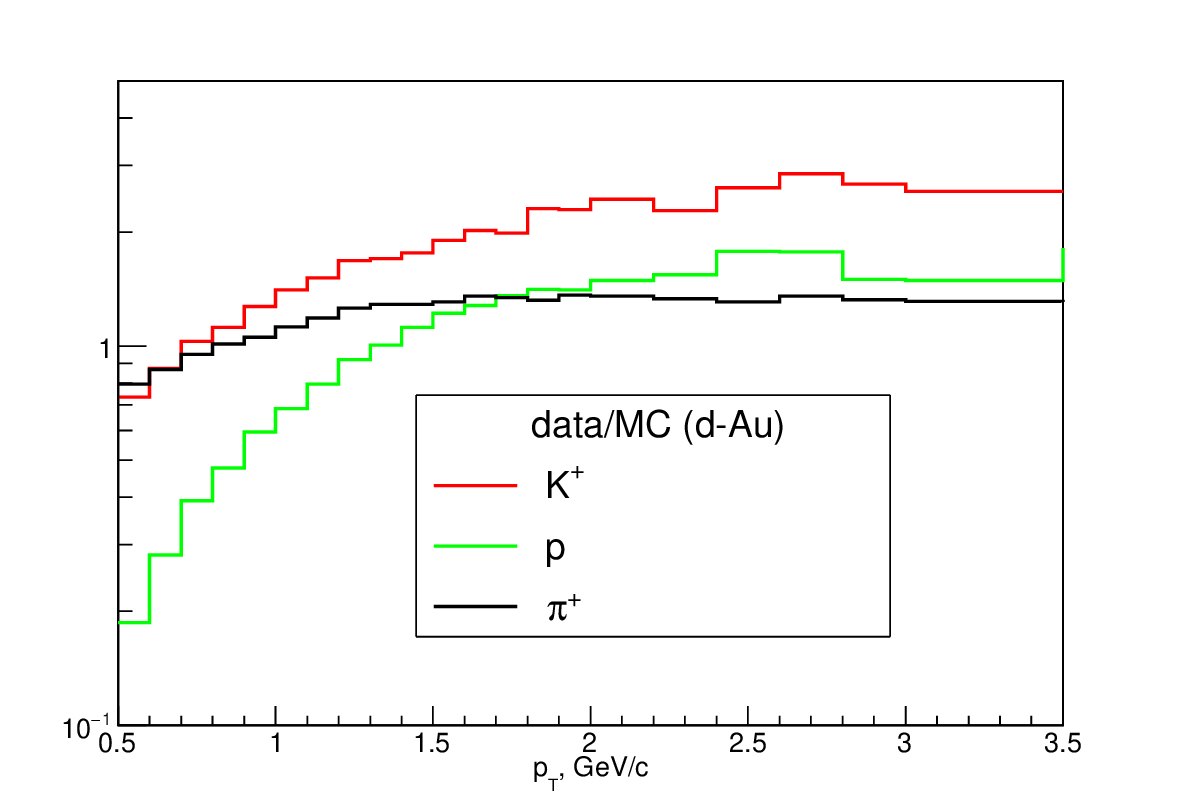}
\caption{\label{fig:3} The Data/MC for d-Au collisions at 200 GeV. }
\end{figure}

\begin{figure*}
\includegraphics[width=1\textwidth]{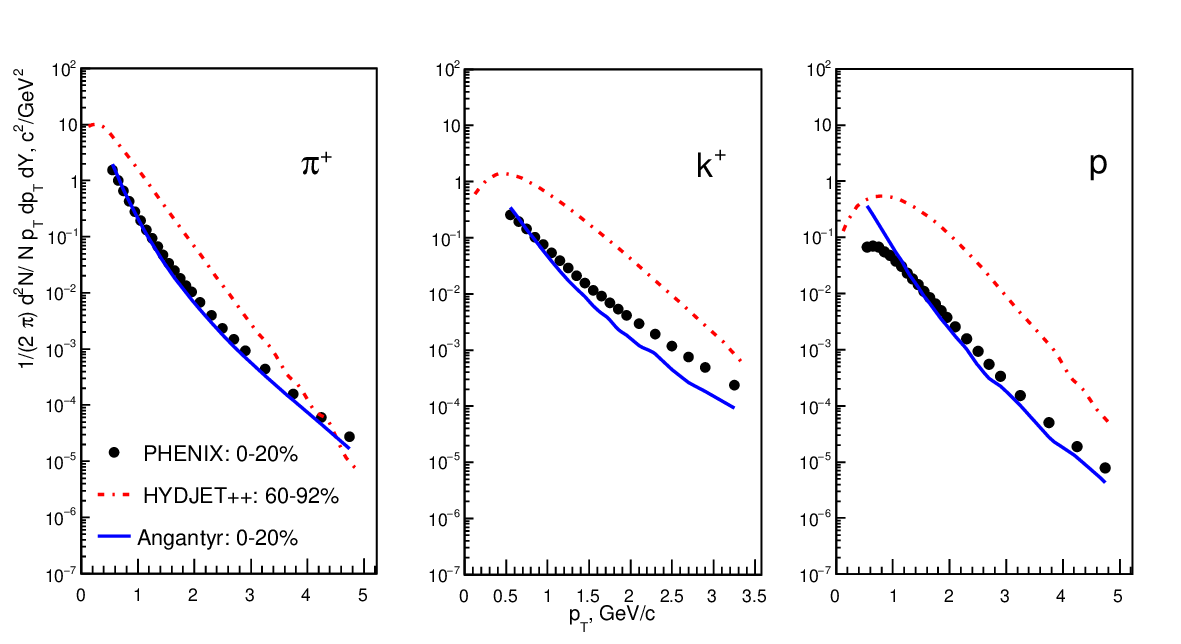}
\caption{\label{fig:3} The dotted lines represent transverse momentum distribution of pions, kaons, and protons in Au-Au collisions at 200 GeV (b=1.92 R ) using HYDJET++ model. The data points are for central d-Au collisions. The solid line represents Angantyr results for central d-Au collisions.}
\end{figure*}

\section{RESULTS AND DISCUSSIONS}

\subsection{Multiplicity and Yields} 

We have simulated 50 million minimum-biased d-Au events with the spatially constrained CR. The charged particle multiplicity in different events in the pseudo-rapidity interval $-3.8<\eta<-2.8$ (Au direction), along with the data from the STAR experiment \cite{STAR:2006kxj}, is shown in Fig. 2. The newly added SR CR mechanism reduces the particle production in central collisions and perfectly reproduces the high-multiplicity tail without further tuning. The $N_{ch}$ distribution of the default model is shown in Fig. 2 for reference. 

Traditionally, the theoretical models assign the centrality based on the impact parameter of the collision that produces a given number of participant nucleons $<N_{part}>$. However, it was observed in p-Pb collision that the $<N_{part}>$ is highly dependent on the type of Glauber model used \cite{Bierlich:2018xfw,ATLAS:2015hkr}. A similar problem is expected in d-Au collision; hence, we will assign centrality based on the final state hadrons produced in the collisions, as done in the experiments, rather than the number of wounded nucleons. The centrality of a collision is defined as the percentile of the hadronic cross section corresponding to a certain multiplicity or energy. It can be calculated from final state multiplicity as follows:\\
1. Let N is the number of events corresponding to a certain $N_{ch}$, and $N_{events}$ is the total number of events. \\
2. The probability of a event producing $N_{lim}$ hadrons is given by:\\
 \begin{equation*}
    P(N_{ch})=\frac{1}{N_{events}}\frac{dN}{dN_{ch}}
\end{equation*}
3. The centrality a collision corresponding to multiplicity $N_{lim}$ is given by:\\

\begin{equation*}
    c(N_{lim})=\int_{N_{lim}}^{\infty} P(N_{ch}) \,dN_{ch}
\end{equation*}
A similar procedure can be carried out with $\Sigma E_T$ of charged hadrons to determine centrality.  In this work, we have assigned centrality based on the charged hadron multiplicity in the pseudo-rapidity interval $-3.8<\eta<-2.8$. The $N_{ch}$ values for each centrality in Pseudo-rapidity interval $-3.8<\eta<-2.8$ are given in Table 1

In Fig. 3, we present the pseudorapidity distribution of charged hadrons in minimum-bias d-Au collisions along with the data from the PHOBOS experiment \cite{PHOBOS:2003fjw}. The Angantyr results show a good agreement with the data in the whole $\eta$ range. The model provides a better description of data than the QGP-based model, like the ``A Multi-Phase Transport Model" (AMPT) \cite{Lin:2003ah} and jet-based model (non-QGP) like HIJING \cite{Gyulassy:1994ew} at backward rapidities.

In Fig. 4, we have shown the transverse momentum distribution of $\pi ^+$, $K^+$, and $p$. The results are shown for three centrality bins: (0-20)\%, (20-40)\%, and (40-60)\%, similar to the PHENIX data \cite{PHENIX:2013kod}. For pions, the data overpredicts our results in central collisions, and the agreement with data improves for the peripheral collisions, eventually matching the data in most peripheral collisions. A similar trend was observed for protons. The model underpredicts the kaon spectra in all centralities.

\begin{table}[h]
\caption{\label{tab:table1}%
The range of $N_{ch}$ for each centrality class.
}
\begin{ruledtabular}
\begin{tabular}{cccc}
\textrm{Centrality}&
\textrm{0-20\%}&
\textrm{20-40\%}&
\textrm{40-60\%}\\
\colrule
$N_{ch}$ & $>17$ & 11-17 & 6-11 \\

\end{tabular}
\end{ruledtabular}
\end{table}

\subsection{In-medium effects}

It is a convention in the heavy ion community to compare the data with a scaled pp event to study the effect of the medium formed. The nuclear modification factor ($R_{AA}$), ``the ratio of the yield of a heavy-ion event and a scaled pp event", is a classic example. The Pythia8/Angantyr provides a good description of a heavy-ion collision without the collective effects. The Angantyr calculations can be used as a model-dependent residual observable to complement quantities like the nuclear modification factor.  

The central Pb-Pb collisions at LHC (2.76 TeV) are most likely to form a QGP. The Pythia8/Angantyr was used to study particle spectra in Pb-Pb collisions at 2.76 TeV in R. Singh et al. \cite{Singh:2021edu,ALICE:2013rdo}. If one observes carefully, the data/MC in ref~[31] and similar papers look surprisingly similar to a typical $R_{AA}$ plot. For example, the magnitude of the data/MC is higher for protons, followed by kaons and pions, similar to $R_{AA}$. The peaks for all particles are at similar positions in both $R_{AA}$ and data/MC. In fact, the data/MC (from Angantyr) are analogous to $R_{AA}$, where the scaled pp data are replaced by Angantyr results. At low and mid $p_T$, the data/MC shows a clear enhancement. Such enhancement can be related to the thermally produced particles; the enhancement is higher for kaons and protons, which can be related to strangeness and baryon enhancement, respectively. In the high $p_T$ regime, a suppression can be seen, which can be associated with jet quenching. A similar pattern can be found in other studies using the model \cite{Bierlich:2022pfr}. The nucleon-nucleon interactions in Angantyr are classified as absorptive (Non-Diffractive), wounded target/projectile, or Secondary Non-Diffractive (treated similarly to Single Diffractive pp events), double diffractive, and elastic sub-collisions. The minimum-bias pp collisions are also separated similarly; however, the cross-sections of the different sub-collisions vary. The cross-sections of the Secondary Non-diffractive (modified Single Diffractive) subcollisions in Angantyr are significantly higher in Angantyr compared to a typical pp collision. The increased Single diffractive sub-collisions in Angantyr reduce the Yield in Angantyr compared to scaled pp data, causing the modification. The modifications due to separations of sub-collisions can be found in Fig. 5. The nuclear modification can either result from the separation of different sub-collisions \cite{Nayak:2024sbp} or due to the presence of a strong collective medium. The distinction between the two types of modifications should be studied to better understand the in-medium effects.

In Fig. 6, we have shown data/MC of pions, kaons, and protons in central d-Au collisions (0-20)\%. For Pions, the data/MC are similar to the Pb-Pb results, but the magnitude of data/MC is noticeably smaller, suggesting a smaller thermal particle production. As discussed earlier, this study was performed using SC CR, which enhances baryon production, leading to a smaller data/MC for protons compared to Pb-Pb collisions. The value of data/MC is still greater than unity in central collisions despite including SC CR. The data/MC is significantly higher for kaons, which can be an indication of strangeness enhancement in central d-Au collisions. Unlike the Pb-Pb case, we did not find a high $p_T$ suppression in data/MC values in d-Au results, indicating the absence of jet quenching in d-Au collisions. In brief, the d-Au collisions have a small thermal particle production and baryon enhancement but no jet quenching.

Thermal model calculations are often used to describe low-$p_T$ particle production in heavy-ion collisions. It would be interesting to see what a typical thermal model will predict for a system with a similar number of participants. The peripheral Au-Au (b=1.92 $R_{Au}$) collisions at 200 GeV have an equal number of participant nucleons as central d-Au collisions.  The HYDJET++ model calculates the particle production on freeze-out hypersurfaces based on preset freeze-out conditions. The model rescales the effective volume of the fireball by a factor of $N_{part}(b)/N_{part}(b=0)$ for the peripheral collisions. We have simulated $5\times 10^5$ events using the HYDJET++ model (only the HYDRO part) for (b=1.92 $R_{Au}$). In Fig. 7, we have shown the HYDJET++ results for peripheral Au-Au collisions and d-Au results from the Angantyr model, along with the data points. The HYDJET++ results significantly overpredict the Angantyr results and the experimental data for all particles. This suggests that the invariant yield observed in the d-Au collisions is smaller than the yield one would expect from a thermal model for a given $N_{part}$. However, the current implementation of PYTHIA can not fully describe the pp and peripheral d-Au data at 200 GeV. Hence, some deviation is expected while using Angantyr results as background.

\section{Summary}
In the current work, we have studied the d-Au collisions at 200 GeV using PYTHIA8/Angantyr. The model simulates d-Au collisions based on the interaction of wounded nucleons without assuming the formation of quark-gluon plasma. We have used a spatially constrained color reconnection, which performs color reconnection between neighboring dipoles from different sub-collisions within a certain spatial constraint. The Pythia8/Angantyr with the spatially constrained color reconnection provides a useful non-hydrodynamic baseline in d-Au collisions at RHIC energies. It provides a good description of multiplicity and reproduces the high-multiplicity tail reasonably well. It provides a better description of the pseudorapidity distribution of charged hadrons in the whole pseudorapidity interval than models like AMPT and HIJING. The Angantyr calculations of $p_T$ spectra of pions and protons agree well with experimental data in peripheral collisions but underpredict the data in central collisions. The model underpredicts the kaon spectra in all centralities.

The data/MC from Angantyr follow similar trends as the nuclear modification factor $R_{AA}$ in Pb-Pb collisions at LHC energies. The trends like mid-$p_T$ peak and high-$p_T$ suppression were observed for all particles, similar to $R_{AA}$. It can be used as a model-dependent observable to complement $R_{AA}$. It can be used as a reference where experimental data is unavailable. The data/MC in the d-Au collisions show similar trends as Pb-Pb collisions, like mid-$p_T$ enhancement of protons and kaons. However, it does not show any suppression at high-$p_T$. However, no high-$p_T$ suppression was observed. Furthermore, the experimental data significantly underpredicts the thermal model result for a similar $N_{part}$.

\begin{acknowledgments}
BKS sincerely acknowledges financial support from the Institution of Eminence (IoE), BHU Grant number 6031. SRN acknowledges the financial support from the UGC Non-NET fellowship and IoE research incentive during the research work. AD acknowledges the institutional fellowship from IIIT DM Jabalpur.  

\end{acknowledgments}

\end{document}